\documentclass[aps,twocolumn,footnote]{revtex4}
\usepackage{blindtext}
\usepackage{graphicx}
\usepackage{xcolor}
\usepackage{subfig}
\usepackage{amssymb}
\usepackage{lipsum}
\usepackage{physics}
\usepackage{mathrsfs}
\usepackage{MnSymbol}
\usepackage{tensor}
\graphicspath{ {images/} }
\usepackage[utf8]{inputenc}
\usepackage{amsmath}
\usepackage{enumerate}
\usepackage{tabularx}
\usepackage{float} 
\usepackage[colorlinks=true, pdfstartview=FitV, linkcolor=blue, citecolor=red, urlcolor=black, breaklinks=true]{hyperref}

\DeclareMathOperator{\arctanh}{arctanh}

\newcommand{\be}{\begin{equation}}
\newcommand{\ee}{\end{equation}}

\newcommand{\PR}[1]{\ensuremath{\left[#1\right]}}
\newcommand{\PC}[1]{\ensuremath{\left(#1\right)}}

\begin{document}
\title{Kink Crystal}
\author{D. Bazeia}
\author{G. S. Santiago}
\affiliation{Departamento de Física, Universidade Federal da Paraíba, João Pessoa, Paraíba, Brazil}

\begin{abstract}
We describe a one-dimensional kink crystal, which represents a collection of equal and equally localized kinks forming a lattice in the real axis. The results are analytical, original and may motivate other studies on localized structures in high energy physics.
\end{abstract}
\maketitle

The study of localized structures in the real line has a very long history. In high energy physics, it started more than 50 years ago, with the unveiling of kinks; see, e.g., Refs. \cite{Enz,Fin,DHN,Jac,Raj,MSu} and references therein. The kink solution is in general a simple analytical configuration that has the form of the hyperbolic tangent. It finds distinct applications in high energy physics \cite{Raj,MSu}, and also appears in condensed matter, where it is used to model domain wall. In magnetic systems, for instance, kinks or domain walls may describe boundaries between regions having distinct magnetic domains \cite{HSc,OHa}. 

After the finding of kinks in high energy physics, researchers have investigated the possibility of constructing lattices or chains of such localized structures. In Ref. \cite{Gri}, for instance, the authors considered thermal creation of kink-antikink pairs in the two-dimensional $\phi^4$ theory in the presence of spontaneous symmetry breaking. This was soon followed by another work, \cite{Man}, in which an array of kink and antikink was studied on the circle. The search for kink or domain wall lattices was also analyzed more recently in \cite{Vacha1,Vacha2}, in particular, in \cite{Vacha1} the authors constructed lattices with alternating kinks and antikinks, which could be stable in certain models. The problem concerning chains alternating kinks and antikinks was further considered very recently in Ref. \cite{Man2}, adding other results. Due to the complicated task, some investigations dealing with lattice or chain of kinks are implemented numerically, while analytical results are mainly obtained when kink and antikink alternate to provide solutions, which may not be stable since kinks and antikinks in general conspire against each other.

There are other well distinct routes to construct lattices of kinks or domain walls. One important possibility concerns investigations of inhomogeneous chiral phase in the quark matter \cite{Rev}. Interesting studies related to this appeared before in Refs. \cite{Dun1,Dun2,Dun3}, where twisted kink crystal is investigated in the $1+1$ dimensional Nambu--Jona-Lasinio or chiral Gross–Neveu model. Another possibility of current interest concerns the presence of kink crystalline condensate, multikink and twisted kink crystal in holographic superconductor; see \cite{Mat1,Mat2} and references therein. In these systems, the presence of periodic inhomogeneous configurations of the kink type is of interest in high energy and in condensed matter physics, with the investigations connecting systems described by the Nambu--Jona-Lasinio, the Gross-Neveu, the Ginzburg-Landau and the nonlinear Schrödinger equations. In the present work, however, the kink crystal which we shall introduce follows another approach, leading to analytical and original solution, different from the kink crystals found before in the related literature.

To investigate the subject, we introduce a model described by two real scalar fields in $1+1$ spacetime dimensions. It is inspired by Refs. \cite{BLM,BMM}: in \cite{BLM} one includes a modification in the kinematics of one of the two fields, which induces  novel profile to the kink; and in \cite{BMM} it is shown how the behavior of fermions in the new kinklike background works, contributing to the presence of new states inside the fermionic gap, besides the fermion zero mode uncovered long ago in \cite{JR}. We start writing the Lagrangian density that defines the system 
\begin{equation}\label{modify}
    \mathcal{L} = \frac{1}{2}f(\chi)\partial_{\mu}\phi\partial^{\mu}\phi + \frac{1}{2}\partial_{\mu}\chi\partial^{\mu}\chi - V(\phi, \chi),
\end{equation}
 where $f(\chi)$ is a non-negative function of the field $\chi$ which modifies the kinematics of the field $\phi$. We consider standard Minkowski metric, natural units, and dimensionless fields and spacetime coordinates. The equations of motion for $\phi(x,t)$ and $\chi(x,t)$ are 
\begin{equation}
\label{EoMtp}
 \partial_{\mu}\PR{f(\chi)\partial^{\mu}\phi} + V_{\phi} = 0,
\end{equation}
and
\begin{equation}
\label{EoMtc}
\partial_{\mu}\partial^{\mu}\chi + V_{\chi} - \frac{1}{2}\frac{d f(\chi)}{d \chi}\partial_{\mu}\phi\partial^{\mu}\phi = 0.
\end{equation}
In the search for static configurations, they can be written in the form
\begin{align}
\begin{split}
\label{eom}
 \frac{d}{dx}\PC{f(\chi)\frac{d\phi}{dx}} = V_{\phi},
\;\;\;\;\;\;\;\;
\frac{d^2\chi}{dx^2} - \frac{1}{2}\frac{df(\chi)}{d\chi}\left({\frac{d\phi}{dx}}\right)^2 = V_\chi,
\end{split}
\end{align}
where $V_\phi={\partial V}/\partial\phi$ and $V_\chi=\partial V/\partial\chi$.

In the above model, the energy density $\rho(x)$ corresponding to static fields $\phi(x)$ and $\chi(x)$ is given by 
\begin{equation}
\label{ed}
    \rho(x)=\frac{1}{2}f(\chi)\PC{\frac{d\phi}{dx}}^2 +\frac{1}{2}\PC{\frac{d\chi}{dx}}^2 + V(\phi,\chi).
\end{equation}
It is possible to introduce a new function $W(\phi,\chi)$ in order to rewrite the above expression as
\begin{eqnarray}
\label{ed1}
    \rho(x) &=& \frac{1}{2}f(\chi)\PC{\frac{d\phi}{dx}-\frac{W_\phi}{f(\chi)}}^2 +\frac{1}{2}\PC{\frac{d\chi}{dx}-W_\chi}^2 \!+\nonumber
    \\
    &&+\,V(\phi,\chi)-
    \frac{1}{2}\frac{W_{\phi}^2}{f(\chi)} - \frac{1}{2}W_{\chi}^2+\frac{dW}{dx},
\end{eqnarray}
where $W_\phi={\partial W}/\partial\phi$ and $W_\chi=\partial W/\partial\chi$. We then see that solutions of the first-order equations 
\begin{align}
\begin{split}
\label{1o}
 \frac{d\phi}{dx} = \frac{W_{\phi}}{f(\chi)},
\;\;\;\;\;\;\;\;
\frac{d\chi}{dx} = W_{\chi},
\end{split}
\end{align}
minimize the energy density when the potential takes the form
\begin{equation}
\label{pmd}
    V(\phi,\chi) = \frac{1}{2}\frac{W_{\phi}^2}{f(\chi)} + \frac{1}{2}W_{\chi}^2.
\end{equation}
One notices that, since $f(\chi)$ is a non-negative function, the potential is bounded from below, which is important to control the physics of the system. Moreover, the minimum energy is obtained in the form
\begin{align}
E_{B}\!=\! \Delta W\!=\! \vert W(\phi (\infty), \chi (\infty))\! - W(\phi(-\infty), \chi (-\infty))\vert.
\end{align}
Interestingly, the energy of the static solutions only depend on their asymptotic values. The above framework is inspired by an old work \cite{Bog}, and is known as the Bogomol'nyi procedure. Solutions that obey the first order equations also solve the equations of motion when $W_{\chi\phi}=W_{\phi\chi}$. They are also stable, since they are minimum energy solutions.

\begin{figure}[ht!]
    \centering{{\includegraphics[width=7cm]{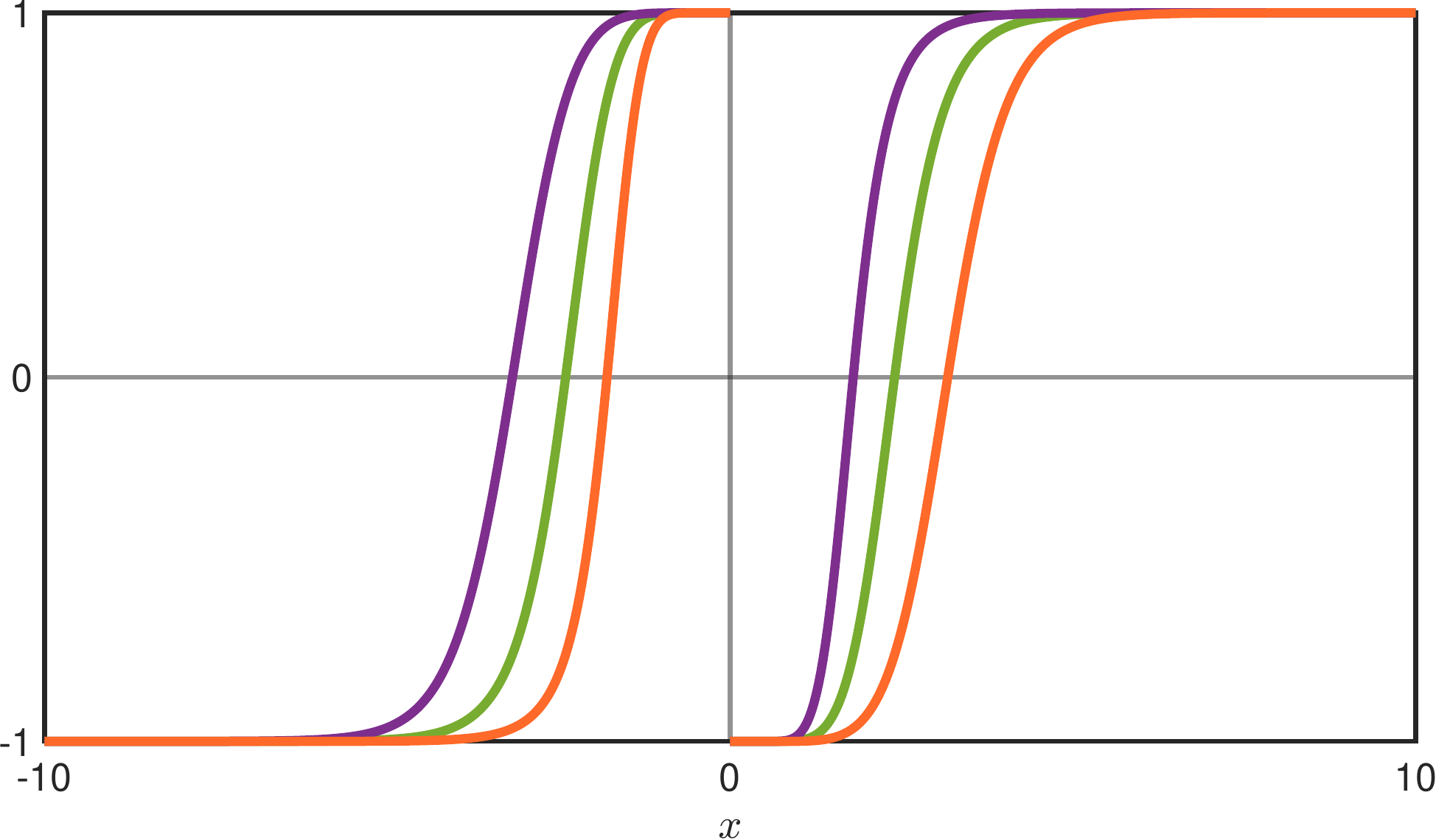}}}%
\caption{The solutions $\phi(x)$ shown in Eq. \eqref{phia} for ${\bar x}_0=-1 \;({\rm orange}), \;0\; ({\rm green})\; {\rm and} \;1\; ({\rm purple})$. }\label{fig1}
\end{figure}

Let us now choose
\begin{equation}
\label{W44}
    W(\phi,\chi) = b^2\phi - \frac{1}{3}\phi^3 + \alpha a^2\chi - \frac{1}{3}\alpha\chi^3,
\end{equation}
where $a$, $b$ and $\alpha$ are positive real constants. Differently from \cite{BLM}, we now take $f(\chi) = \chi^2$, which leads to the new potential
\begin{equation}
    V(\phi,\chi) = \frac{1}{2}\frac{(b^2-\phi^2)^2}{\chi^2} + \frac{1}{2}\alpha^2(a^2-\chi^2)^2.
\end{equation}
The presence of $\chi^2$ in the denominator of the first term of the potential induces a divergence at $\chi = 0$, but this is precisely what makes the model interesting. To see this, one uses \eqref{1o} to write the first-order equations
\begin{align}
\begin{split}
\frac{d \phi}{dx} = \frac{b^2-\phi^2}{\chi^2},
\;\;\;\;\;\;\;\;
\frac{d \chi}{dx} = \alpha(a^2-\chi^2).
\end{split}
\end{align}
The equation for the field $\chi$ above can be solved independently. The solution can be written as
\begin{equation}
\label{ctan}
    \chi (x)=a\tanh\PC{\alpha a (x - x_{0})},
\end{equation}
where $x_{0}$ is a constant of integration which is related to the center of the solution, and $a$ is the corresponding amplitude. For simplicity, we take $a=1$ and $x_{0} = 0$, indicating that the center of the $\chi(x)$ configuration is at the origin $x=0$. Substituting this result in the first-order equation for $\phi$, we get the solution
\begin{equation}\label{phia}
    \phi(x) = b\tanh\left(\frac{b}{\alpha}\PC{\alpha x - \coth(\alpha  x)} + {\bar x}_0\right),
 \end{equation}   
 where ${\bar x_0}$  is another integration constant. However, compared to the integration constant of the $\chi$ field, this new constant has a different meaning: it induces a modification in the form of the solution, as it is illustrated in Fig. \ref{fig1} for $b=1$. Although this may be of interest, we shall not explore the effects of the integration constant ${\bar x}_0$ in this work. Our main motivation here is to construct a kink crystal, and we shall do it considering the simplest possibility. Other effects related to the amplitude and integration constants can be easily incorporated in future investigations. 

In the model investigated above, we draw attention to two distinct features. The first one is that the discontinuity that appears at the center of the $\phi(x)$ solution is a consequence of the term $1/\chi^2$ in the potential; however, it causes no problem since the energy density (see below) is well-defined and integrates to a finite value. The second feature is that the solution now engenders the form of a kink-kink configuration, with the center of both kinks and their relative location being controlled by the parameter $\alpha$, as we further describe below. 

\begin{figure}[ht!]
    \centering{{\includegraphics[width=9cm]{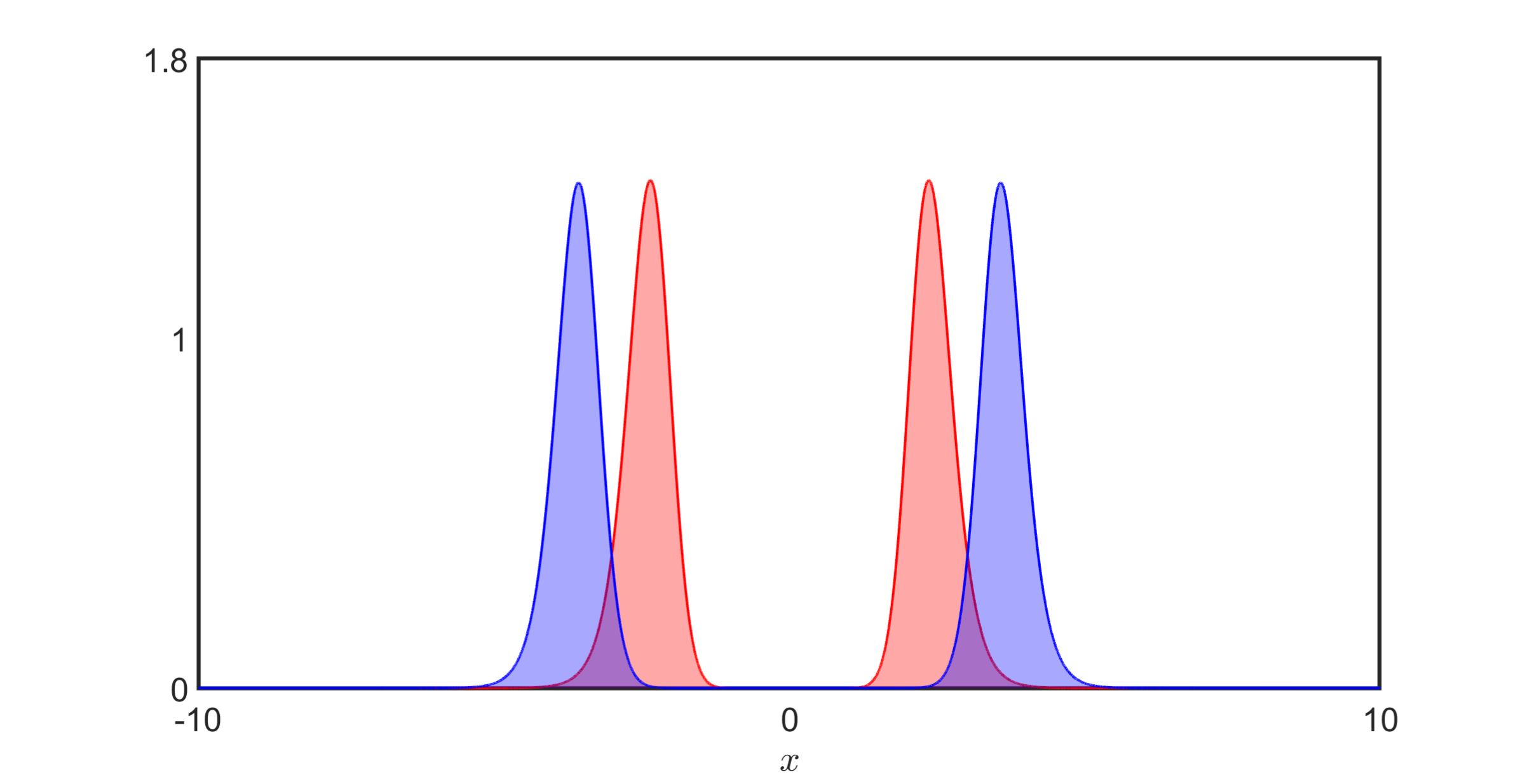}}}%
\caption{The energy density $\rho_{1}(x)$, depicted for $\alpha=1/3$ (blue) and $\alpha=1/2$ (red). }\label{fig2}
\end{figure}

It is possible to use Eq. \eqref{ed} to write the energy density for the above solutions in the form $\rho = \rho_{1} + \rho_{2}$, where 
\begin{equation}
    \rho_{1}(x) = \sech^4 \PC{  x - \frac{1}{\alpha} \coth(\alpha x)}\coth^2(\alpha x),\end{equation}
and $\rho_{2}(x) =  \alpha^2\sech^4(\alpha x)$.
Since the $\rho_2$ term only depends of the $\chi$ field, which is described by the well-known $\chi^4$ model, we only depict the $\rho_1$ contribution in Fig. \ref{fig2}. There we are using $\alpha=1/3$ and $\alpha=1/2$, and we see two distinct lump-like energy densities, associated to the kink-kink configuration illustrated in Fig. \ref{fig1}. 

One notices that the maxima of the energy density occur exactly at the center of each kink in the kink-kink solution.

\begin{figure}[ht!]
    \centering{{\includegraphics[width=7cm]{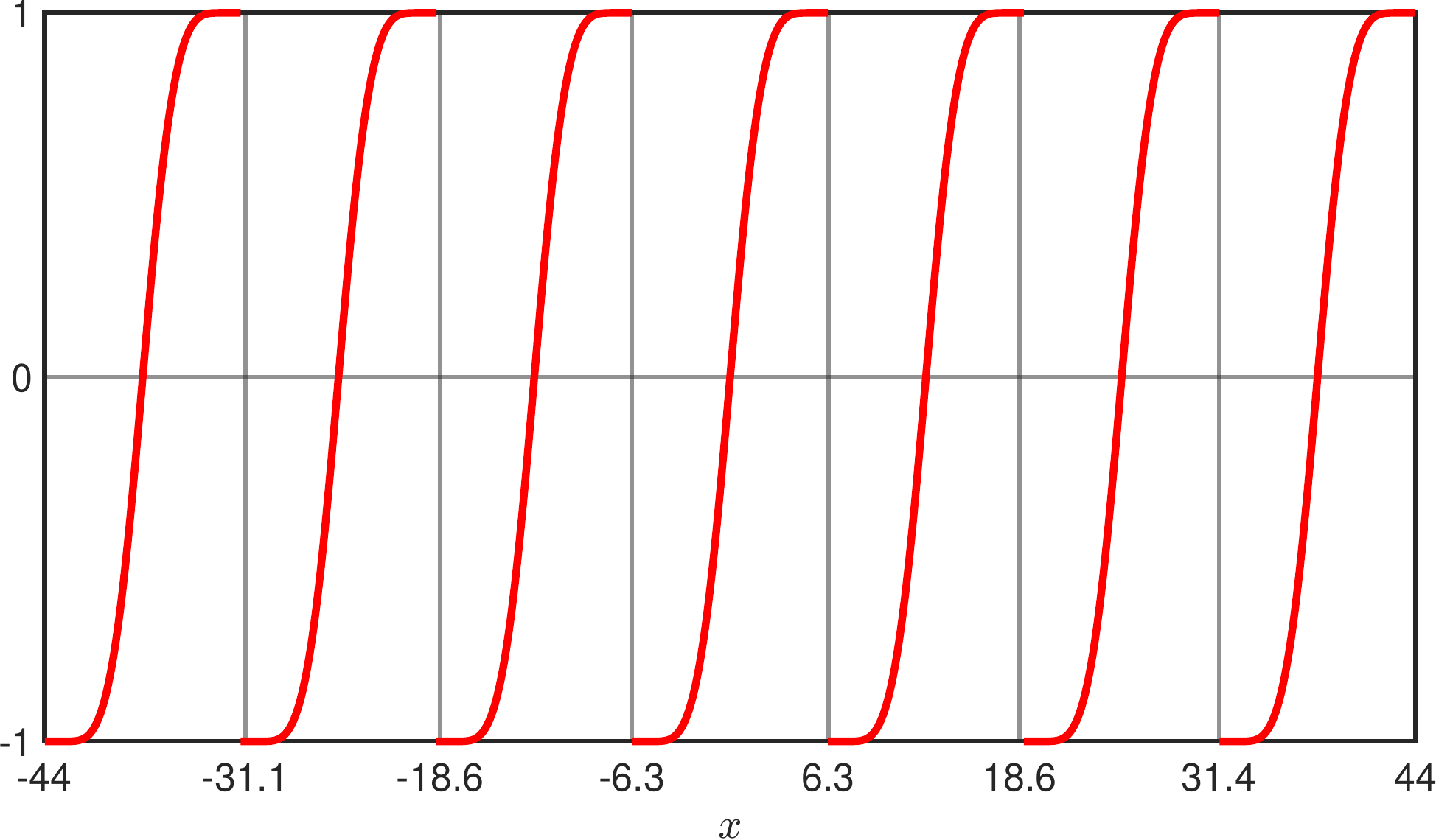}}}%
\caption{Solution of the kink crystal, using $\alpha=1/2$.}\label{fig3}
\end{figure}

Since the function $f(\chi)=\chi^2$ was able to split the $\phi(x)$ kink into two distinct kinks in the kink-kink configuration, we then think of finding another function, capable of generating a lattice of kinks. This would be the kink crystal which we are searching for, inspired by the comments introduced in the beginning of this work. After some calculations, we have found the interesting function
\begin{equation}
    f(\chi)=1+\cos(\arctanh(\chi)).
    \end{equation}
In this case, using \eqref{W44} the potential can be written as
\begin{equation}
    V(\phi,\chi) = \frac{1}{2}\,\frac{(1-\phi^2)^2}{1+\cos\PC{\arctanh({\chi})}} + \frac{1}{2}\,\alpha^2(1-\chi^2)^2.
\end{equation}
There are divergences whenever $\arctanh(\chi) = (2n+1)\pi$, with $n\in \mathbb{Z}$, but this cause no problem, as we further explain below. Indeed, the model still attains first-order equations in the form 
\begin{align}\label{foecry}
\begin{split}
\frac{d \phi}{dx} = \frac{1-\phi^2}{1+\cos\PC{\arctanh(\chi})},
\;\;\;\;\;\;\;\;
\frac{d \chi}{dx} = \alpha\,(1-\chi^2).
\end{split}
\end{align}
The first-order equation for the field $\chi$ was already solved above, so we use this result on the first-order equation for $\phi$ to get
\begin{equation}
\label{phicry}    \phi(x) = \tanh\PC{\frac{1}{\alpha}
    \tan\PC{\frac{\alpha  x}{2}}},
\end{equation}
where we have also discarded an integration constant, since it affects the solution in a way similar to the case displayed in Fig. \ref{fig1}. The above solution is depicted in Fig. \ref{fig3} for $\alpha=1/2$. It represents the kink crystal.

It is important to notice that $\phi(x)$ exhibit a lattice behavior, where multiple equal and equally spaced kinks are formed, in the form of a ...kink-kink-kink-kink... configuration, and not in the form of ...kink-antikink-kink-antikink..., which appeared before in several investigations. The kink crystal is a direct consequence of the periodicity of the function $f(\chi)$, and one further observes that the zeros of $f(\chi)$, when calculated for the solution $\chi(x)= \tanh(\alpha x)$ are compensated by the zeros of $1-\phi^2$, which appear in the potential and in the first order equations, for the solution $\phi(x)$ shown in Eq. \eqref{phicry}. Indeed, $1-\phi^2$ goes to zero faster than $f(\chi)$ at the solutions. Thus, the model behaves adequately, in the presence of the above field configurations.

\begin{figure}[ht!]
    \centering{{\includegraphics[width=8cm]{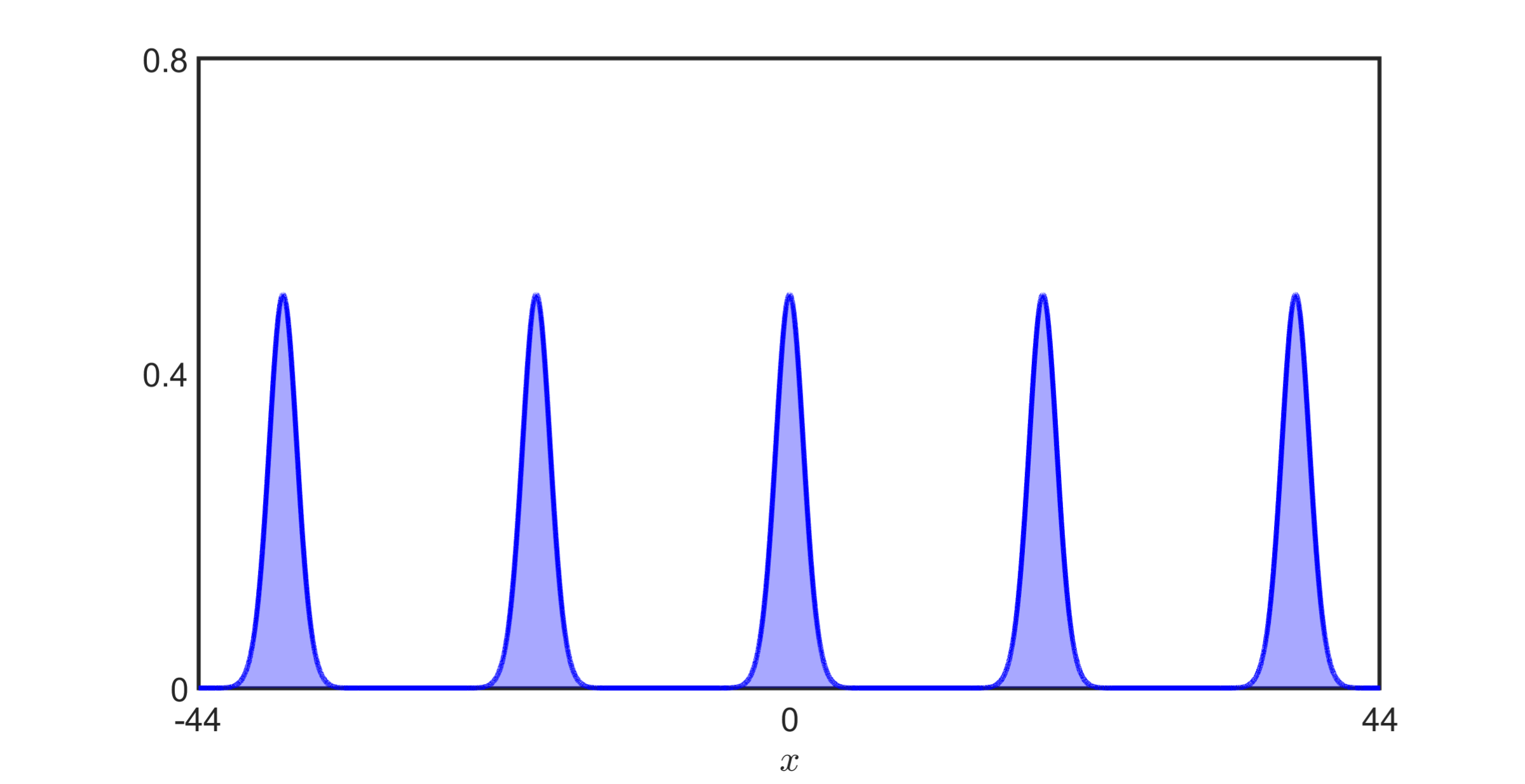}}}
    \centering{{\includegraphics[width=8cm]{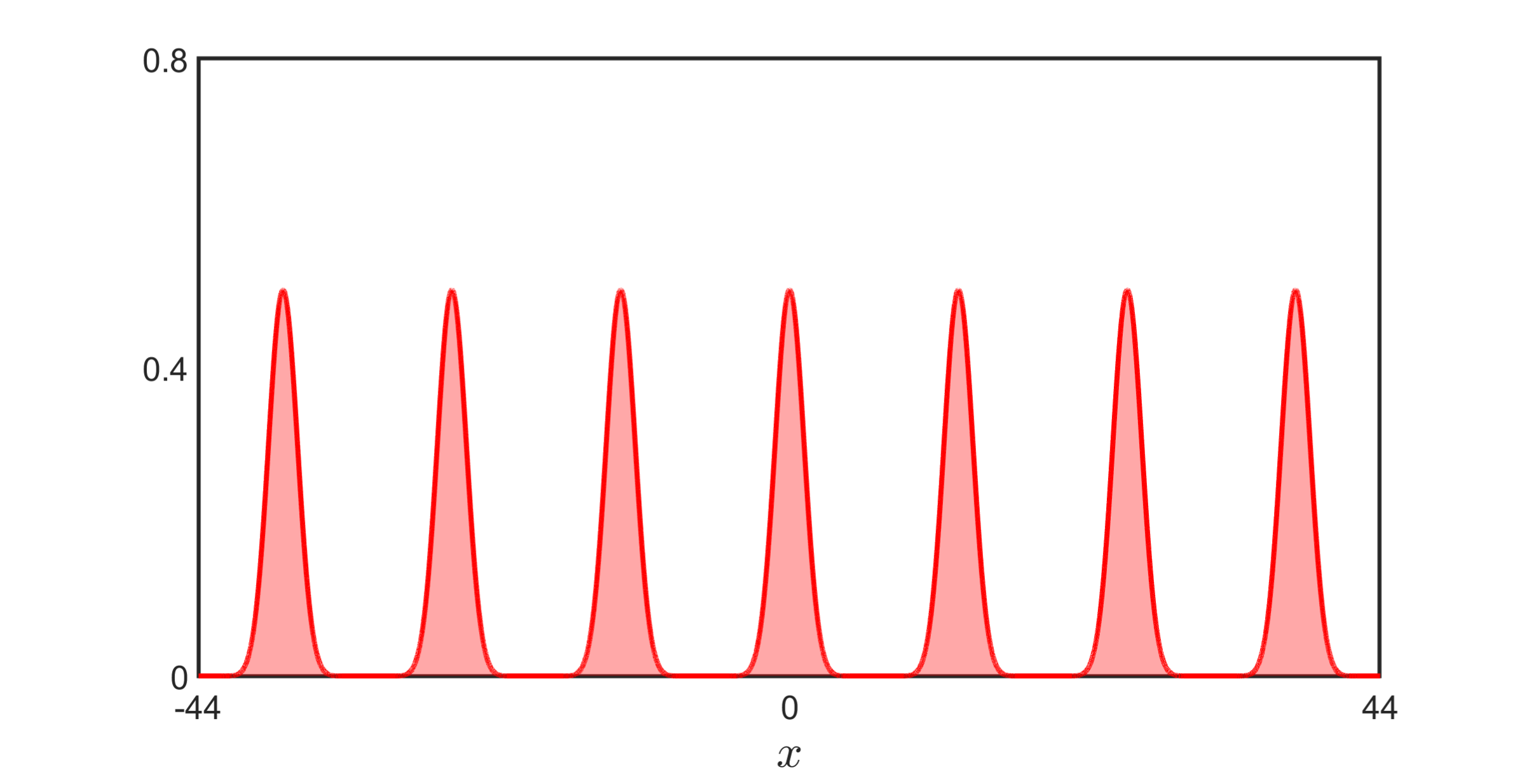}}}
\caption{Energy density of the kink crystal, using  $\alpha=1/3$ (top panel, blue) and $\alpha= 1/2$ (bottom panel, red). }\label{fig4}
\end{figure}

We can verify from the $\phi(x)$ configuration, that the center of the kinks are spaced by $d=2\pi/\alpha$. 
Moreover, using Eq. \eqref{ed}, the energy density can also be rewritten as
$\rho=\rho_1+\rho_2$, with $\rho_1$ now given by
\begin{equation}
\rho_{1}(x) = \sech^4\PC{\frac{1}{\alpha}\tan\PC{\frac{\alpha x}{2}}}\PC{\frac{1}{1+\cos\PC{\alpha x}}} ,
\end{equation}
which we depict in Fig. \ref{fig4} for $\alpha=1/3$ and $\alpha=1/2$. We notice that $\alpha$ does not modify the height of the maxima, but it controls the distance between the kinks, that is, it controls the lattice spacing, being related to the lattice constant $d=2\pi/\alpha$ that specify the kink crystal. This can also be visually identified in Fig. \ref{fig4}. We can also calculate the energy per kink in the kink crystal; it gives the value $E=4/3$.

\begin{figure}
    \centering{{\includegraphics[width=7.2cm]{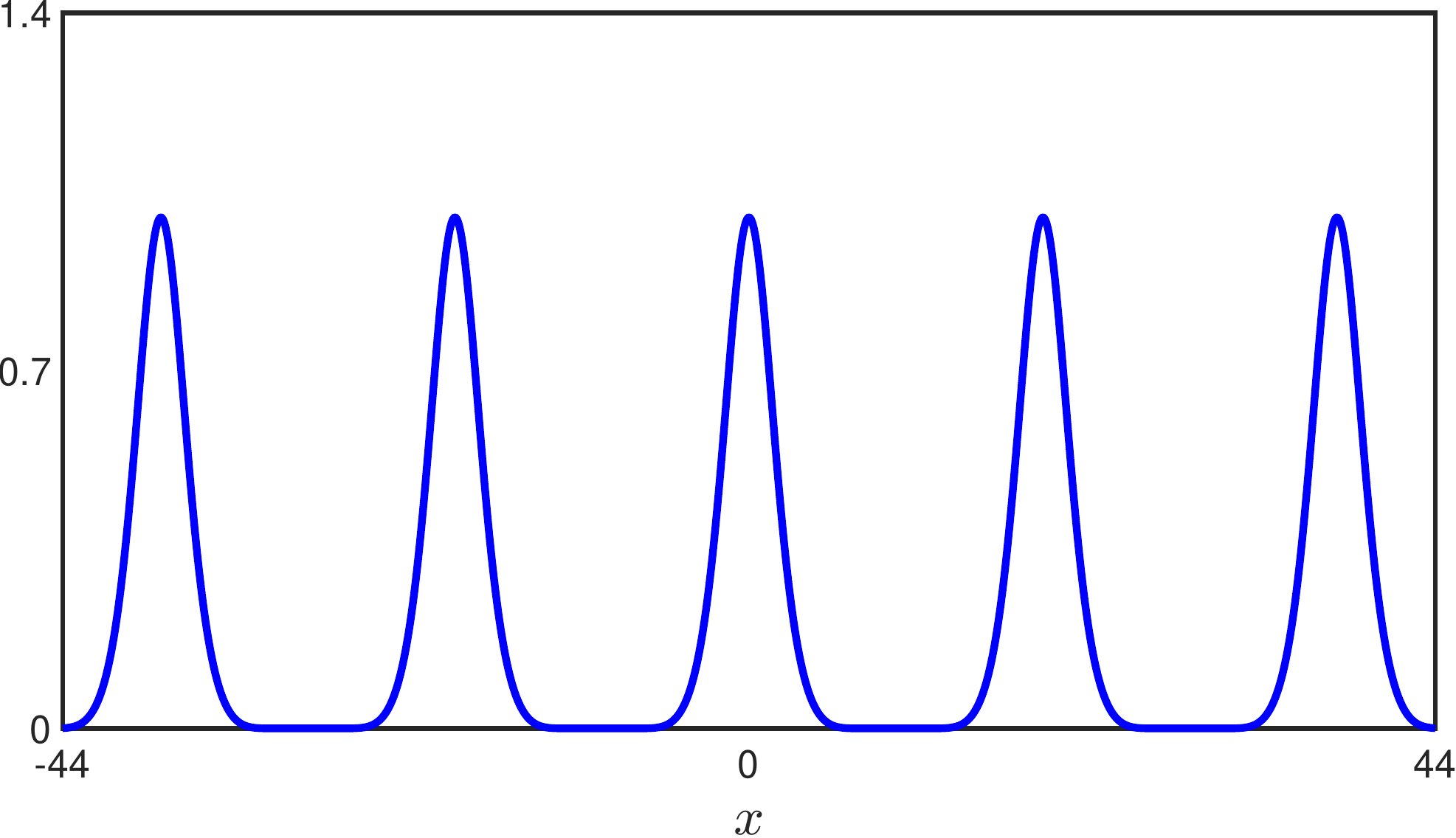}}}
    \centering{{\includegraphics[width=7.2cm]{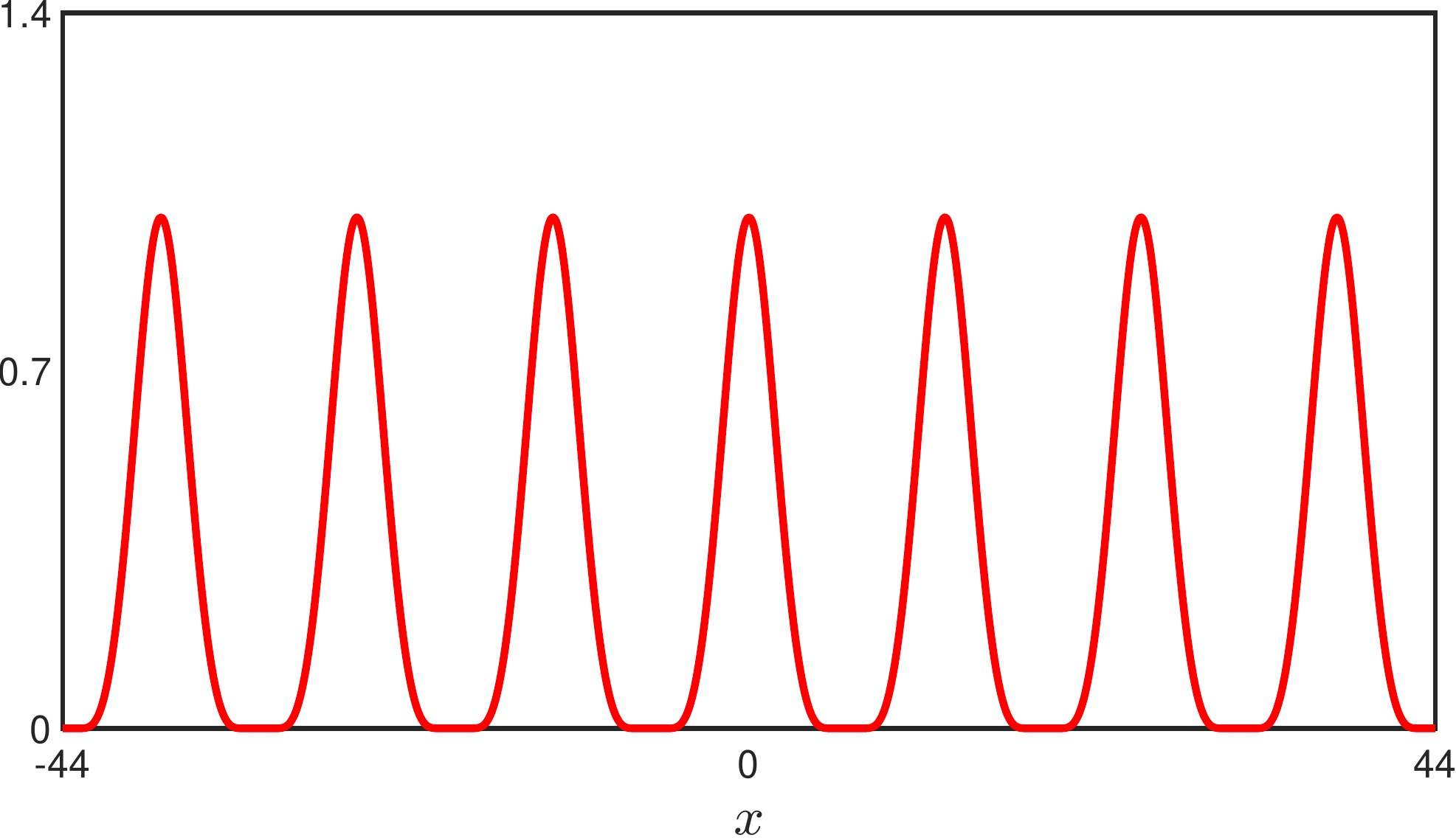}}}
\caption{Behavior of the zero mode, for $\alpha=1/3$ (top panel, blue) and $\alpha=1/2$ (bottom panel, red).}\label{fig5}
\end{figure}

 Although the kink crystal is stable configuration, since it fulfills the Bogomol'nyi bound, it is also of interest to study linear stability and the behaviour of the zero mode. This can be done considering small perturbations around the static solutions, in the form of $\phi(x,t) = \phi(x) + \eta(x)\cos(\omega t)$ and $\chi(x,t) = \chi(x) + \zeta(x)\cos(\omega t)$, with $\eta$ and $\zeta$ as small fluctuations. Substituting these expressions on the equations of motion \eqref{EoMtp} and \eqref{EoMtc} we get, up to first order in $\eta$ and $\zeta$,
\begin{equation}
\begin{split}
-\eta_{xx} -&\frac{1}{f}f_{\chi}\chi_{x}\eta_{x} + \frac{1}{f}V_{\phi \phi}\eta - \frac{1}{f}f_{\chi}\phi_{x}\zeta_{x} +
\\+& \frac{1}{f}\PC{V_{\phi\chi} - \phi_{x}\chi_{x}f_{
\chi\chi}-\phi_{xx}f_{\chi}}\zeta = \omega^2 \eta ,
\end{split}
\end{equation}
and
\begin{equation}
\begin{split}
-\zeta_{xx} + \PC{V_{\chi\chi} + \frac{1}{2}f_{\chi\chi}\phi_{x}^2}\zeta + f_{\chi}\phi_{x}\eta_{x} + V_{\chi\phi}\eta = \omega^2 \zeta.
\end{split}
\end{equation}
Despite the difficulty, but since the static solutions obey the first-order equations \eqref{foecry}, we have been able to prove that $\eta_0(x)\propto d\phi(x)/dx$ and $\zeta_0(x)\propto d\chi(x)/dx$ solve the above equations for $\omega=0$. Therefore, using \eqref{phicry}, we have that $\eta_0(x)$ is proportional to
\begin{equation}
    \sech^2\PC{\frac{1}{\alpha}
    \tan\PC{\frac{\alpha  x}{2}}} \sec^2\PC{\frac{\alpha x}{2}},
\end{equation}
which has no node per period, as we display in Fig. \ref{fig5}.

When we search the literature for chains of kinks, an important and distinct quality of the above model is that it supports a lattice of kinks, which is stable, analytical and original. The result is of current interest, and engenders several distinct directions of investigation,
in particular, the inclusion of fermions, as considered in \cite{BMM}, to see how they behave in the background of the kink crystal. We can also consider extensions of the present investigation to two spatial dimensions, to investigate crystals in the plane, in particular, the case of vortices. Studies in the suggested directions are presently under consideration, and we hope to return to them in the near future. 

{{\bf Acknowledgments:} This work is partially financed by Conselho Nacional de Desenvolvimento Científico e Tecnológico (CNPq), Grant 303469/2019-6 (DB), Coordenação de Aperfeiçoamento de Pessoal de Nível Superior (CAPES), Grant 88887.899555/2023-00 (GSS),  and by  Paraiba State Research Foundation, Grant 0015/2019.}

{{\bf Conflict of interest:} The authors declare that they have no known competing financial interests or personal relationships that could have
appeared to influence the work reported in this paper.}

\end{document}